\documentclass{aa}
\begin{document}
\title{On  the nature of the GRB-SGRs blazing jet }
\author{D. Fargion}
\institute{Physics Dept. Rome, Univ. 1; INFN, Rome; Ple A. Moro
2.Italy}
\date{Received / Accepted}
\maketitle

\begin{abstract}
Gamma Ray Burst and Soft Gamma Repeaters are neither standard
candle nor isotropic explosions. Our model explain them as strong
blazing of a light-house, spinning and  precessing gamma jet. Such
jets at maximal output ( as  GRBs in Supernova like sources at
cosmic edges) or at late lower power stages (SGRs in nearer
planetary nebulae in galactic halo) may blaze the observer by
extreme beaming $(\Omega < 10^{-8} )$ and apparent huge
luminosity. 
\end{abstract}
\section{Introduction}

Gamma Ray Bursts as recent GRB990123 and  GRB990510 emit, for
isotropic explosions, energies as large as  two solar masses
annihilation. These energies are underestimated because of the
neglected role of comparable ejected MeV (Comptel signal)
neutrinos bursts. These extreme power cannot be explained with any
standard spherically symmetric Fireball model. A too heavy black
hole or Star would be unable to coexist with the shortest
millisecond time structure of Gamma ray Burst. Beaming of the
gamma radiation may overcome the energy puzzle. However any mild
explosive beam $(\Omega > 10^{-2} )$ would not solve the jet
containment at the corresponding disruptive energies. Only extreme
beaming $(\Omega < 10^{-8} )$, by a slow decaying, but long-lived
precessing jet, it may coexist with characteristic Supernova
energies, apparent GRBs output, the puzzling GRB980425 statistics
as well as the GRB connection with older,nearer and weaker SGRs
relics. GRBs were  understood as isotropic Fireball and SGRs are
still described  by isotropic galactic explosions (Magnetars).
However  early and late Jet models (Fargion 1994-1998,Blackmann et
all 1996) for GRBs are getting finally credit. Will be possible to
accept a jet model for GRBs while keeping alive a mini fireball
for SGRs? Indeed the strong SGR events (SGR1900+14, SGR1642-21)
shared the same hard spectra of classical GRBs. One should notice
the GRB-SGR similar spectra, morphology and temporal evolution
within GCN/BATSE trigger 7172 GRB981022 and 7171 GRB981022. Nature
would be perverse to mimic two very comparable events at same
detector, same day, by same energy spectra and comparable time
structures by two totally different processes: a magnetar versus
Jet GRBs. We argue here that, apart of the energetics, both of
them are blazing of powerful jets (NS or BH); the jet are spinning
and precessing source in either binary or in accreting disk
systems. The optical transient OT  of GRB might be the coeval
SN-like explosive birth of the jet related to its maximal
intensity; the OT is absent in older relic Gamma jets, the SGRs.
Their explosive memory  is left around their relic nebula or
plerion  injected by the  Gamma Jet which is running away. The
late GRB OT,days after the burst, is related to the explosion
intensity; it is enhanced only by a partial beaming $(\Omega
\simeq 10^{-2} )$. The extreme peak OT during GRB990123 (at a
million time a Supernova luminosity) is the beamed $(\Omega \leq
10^{-5} )$ Inverse Compton optical tail, responsible of the same
extreme gamma (MeV) extreme beamed $(\Omega \leq 10^{-8})$ signal.
The huge energy bath (for a fireball model) on GRB990123 imply a
corresponding neutrino burst. As in hot universe, if entropy
conservation holds, the energy density factor to be added to the
photon $\gamma$ GRB990123 budget is at least $( \simeq
(21/8)\times (4 /11)^{4/3} )$. If entropy conservation do not hold
the energy needed is at least a factor $[21/8]$ larger than the
gamma one. The consequent total energy-mass needed for the two
cases are respectively 3.5 and 7.2 solar masses. No fireball by NS
may coexist with it. Jet could. Finally Fireballs are unable to
explain the key questions related to the association GRB980425 and
SN1998bw (Galama et all1998): (1) Why nearest ``local'' GRB980425
in ESO 184-G82 galaxy at redshift $z_2 = 0.0083$ and the farest
``cosmic'' GRB971214 (Kulkarni et al.
 1998) at redshift $z_2 = 3.42$ exhibit a huge average and peak
 intrinsic luminosity ratio?
$ \frac{<L_{1 \gamma}>}{<L_{2 \gamma}>}  \cong  \frac{<l_{1
\gamma}>}{<l_{2 \gamma}>} \frac{z_{1 }^2}{z_{2 }^2} \cong 2 \cdot
10^5 \;\;; \left. \frac{L_{1 \gamma}}{L_{2 \gamma}} \right|_{peak}
\simeq 10^7 $. Fluence ratios $E_1 / E_2$ are also extreme $\geq 4
\cdot 10^5$.\\
 (2) Why GRB980425 nearest event spectrum is softer
than cosmic GRB971214 while Hubble expansion would imply the
opposite by a redshift factor $(1+z_1)\sim 4.43$?(3) Why,
GRB980425 time structure is slower and smoother than cosmic
one,against Hubble law? (4) Why we observed so many (even just
one) nearby GRBs? Their probability to occur, with respect to a
cosmic redshift  $z_1 \sim 3.42$ must be suppressed by a severe
volume factor $ \frac{P_{1}}{P_{2}} \cong
\frac{z_{1}^{3}}{z_{2}^{3}} \simeq 7 \cdot 10^{7} $. New GRB
fireballs classes are ad hoc and fine-tuned solutions. We proposed
since 1993 (Fargion 1994) that spectral and time evolution of GRB
are made up blazing beam gamma jet GJ. The GJ is born by ICS of
ultrarelativistic (tens GeV) electrons (pairs) on source IR, or
diffused companion IR, BBR photons (Fargion,Salis 1995-98).The
target thermal photons number density may reach a few hundreds to
billions $cm^{-3}$. The thin beamed electron pair jet produce a
coaxial gamma jet. It solves the GRBs energetic by the geometrical
enhancement . Relativistic kinematic would imply $\theta \sim
\frac{1}{\gamma_e}$, where $\gamma_e$ is found to reach $\gamma_e
\simeq 10^3 \div 10^4$ (Fargion 1994,1998). Unique impulsive GRB
jet burst (Wang \& Wheeler 1998) increases the apparent luminosity
by $\frac{4 \pi}{\theta^2} \sim 10^7 \div 10^9$ but face a severe
obvious probability puzzle.In particular we considered (Fargion
1998) an unique scenario where primordial GRB jets decaying in
hundred and thousand years become the observable nearby SGRs. The
ICS on BBR leads to GRBs spectrum (Fargion,Salis 1995,1996,1998):
$\frac{dN_{1}}{dt_{1}\,d\epsilon_{1}\,d\Omega _{1}}$ is
\begin{equation}
\epsilon _{1}\ln \left[ \frac{1-\exp \left( \frac{-\epsilon
_{1}(1-\beta \cos \theta _{1})}{k_{B}\,T\,(1-\beta )}\right)
}{1-\exp \left( \frac{-\epsilon _{1}(1-\beta \cos \theta
_{1})}{k_{B}\,T\,(1+\beta )}\right) }\right] \left[ 1+\left(
\frac{\cos \theta _{1}-\beta }{1-\beta \cos \theta _{1}}\right)
^{2}\right] \label{eq3}
\end{equation}
scaled by a proportional factor $A_1$ related to the electron jet
intensity. The adimensional photon number rate as a function of
the observational angle $\theta_1$ responsible for peak luminosity
becomes $\approx \theta^{-3}$ . The consequent total fluence at
minimal impact angle $\theta_{1 m}$ responsible for the average
luminosity  is $\simeq (\,\theta _{1m})^{-2}$. Assuming a beam jet
intensity $I_1$ comparable with maximal SN luminosity, $I_1 \simeq
10^{45}\;erg\,s^{-1}$, and replacing this value in adimensional
$A_1$ in equation \ref{eq3} we find a maximal apparent GRB power
for beaming angles $10^{-3} \div 3\times 10^{-5}$, $P \simeq 4 \pi
I_1 \theta^{-2} \simeq 10^{52} \div 10^{55} erg \,s^{-1}$ within
observed ones. We also assume a power law jet time decay as
follows
$
  I_{jet} = I_1 \left(\frac{t}{t_0} \right)^{-\alpha} \simeq
  10^{45} \left(\frac{t}{3 \cdot 10^4 s} \right)^{-1} \; erg \,
  s^{-1}
$
where ($\alpha \simeq 1$) able to reach, at 1000 years time
scales, the present known galactic microjet (as SS433) intensities
powers: $I_{jet} \simeq 10^{38}\;erg\,s^{-1}$. We used the model
to evaluate if April precessing jet might hit us once again.
\section{GRBs 980425-980712 and SGRs-GRBs links}
Therefore the jet model answers to the puzzles (1-4): the
GRB980425 has been observed off-axis by a cone angle wider than
$\frac{1}{\gamma}$ thin jet  by a factor $a_2 \sim 500$ (Fargion
1998),i.e few degree wide; we observed only the peripheral  cone
jet whose spectrum is softer and whose time structure is slower
(larger impact parameter angle) and intensity strongly reduced. A
simple statistics favored a repeater hit. GRB980712 trigger 6917
was within $1.6 \sigma$ angle away from April event. Trigger 6918,
repeated making the combined probability to occur quite rare
($\leq 10^{-3}$). Because the July event has been sharper in times
($\sim 4 \,s $) than the April one ($\sim 20 \,s $), the July
impact angle was smaller:$a_3 \simeq 100$. This value is
compatible with the expected peak-average luminosity flux
evolution : $\frac{L_{04\,\gamma}}{L_{07\,\gamma}} \simeq
\frac{I_2\,\theta_2^{-3}}{ I_3\,\theta_3^{-3}} \simeq \left(
\frac{t_3}{t_2} \right)^{-\alpha} \,\left( \frac{a_2}{a_3}
\right)^{\,3} \leq 3.5$ where $t_3 \sim 78 \; day$ while $t_2 \sim
2 \cdot 10^5 \, s$. The predicted fluence are comparable with the
observed ones $\frac{N_{04}}{N_{07}} \simeq
\frac{<L_{04\,\gamma}>}{<L_{07\,\gamma}>} \, \frac{\Delta
\tau_{04}}{\Delta \tau_{07}} \simeq \left( \frac{t_3}{t_2}
\right)^{-\alpha} \,\left( \frac{a_2}{a_3} \right)^2
\,\frac{\Delta \tau_{04} }{\Delta \tau_{07}} \geq 3$.
 Last year 1998 SGR1900+14 and SGR1627-41  events did
 exhibit at peak intensities spectra comparable
with classical hard GRBs. We  proposed their nature as the late
stages of jets fueled by a disk or a companion (WD,NS) star. Their
binary angular velocity $\omega_b$ reflects the beam evolution
$\theta_1(t) = \sqrt{\theta_{1 m}^2 + (\omega_b t)^2}$ or more
generally a multiprecessing angle $\theta_1(t)$ (Fargion \& Salis
1996) which keeps memory of the pulsar jet spin ($\omega_{psr}$),
precession by the binary $\omega_b$ and additional nutation due to
inertial momentum anisotropies or beam-accretion disk torques
($\omega_N$). The complicated spinning and precessing jet blazing
is responsible for the wide morphology of GRBs and SGRs as well as
their internal periodicity. The different geometrical
observational angle might compensate the April 1998 low peak gamma
luminosity ($10^{-7}$) by a larger impact angle which compensates,
at the same time, its statistical rarity ($\sim 10^{-7}$) its
puzzling softer nature and longer timescales. Such precessing jets
may explain (Fargion \& Salis 1995) the external twin rings around
SN1987A. We predicted its relic jet to be found in the South-East
due to off-axis beaming acceleration. Jets may propel and inflate
plerions as the observed ones near SRG1647-21 and SRG1806-20.
Optical nebula NGC6543 (``Cat Eye'') and its thin jets fingers (as
Eta Carina ones), the  double cones sections in Egg Nebula CRL2688
are the most detailed and spectacular lateral view of such jets.
Their blazing in-axis would appear as SGRs or, at maximal power at
their SN birth, as GRBs.

\end{document}